\documentclass[preprint,aps,pre,showpacs,floatfix]{revtex4}
\usepackage[dvips]{graphicx}
\usepackage{amsmath}
\usepackage{epsfig}

\newcommand{\subheading}[1]{\textbf{#1.}}

\begin{document}
\draft \preprint{draft}
\title{Full reduction of large finite random Ising systems by RSRG.}
\author{Avishay Efrat and Moshe Schwartz}
\address{School of Physics and Astronomy, Tel Aviv University, Ramat Aviv,
         Tel Aviv 69978, Israel}
\date{\today}
\begin{abstract}
We describe how to evaluate approximately various physical
interesting quantities in random Ising systems by direct
renormalization of a finite system. The renormalization procedure
is used to reduce the number of degrees of freedom to a number
that is small enough, enabling direct summing over the surviving
spins. This procedure can be used to obtain averages of functions
of the surviving spins. We show how to evaluate averages that
involve spins that do not survive the renormalization procedure.
We show, for the random field Ising model, how to obtain
$\overline{\Gamma(\overrightarrow{r})} =
\overline{\langle\sigma(0)\sigma(\overrightarrow{r})\rangle -
\langle\sigma(0)\rangle\langle\sigma(\overrightarrow{r})\rangle}$,
the "connected" correlation function and
$\overline{S(\overrightarrow{r})} =
\overline{\langle\sigma(0)\sigma(\overrightarrow{r})\rangle}$, the
"disconnected" correlation function. Consequently, we show how to
obtain the average susceptibility and the average energy. For an
Ising system with random bonds and random fields we show how to
obtain the average specific heat. We conclude by presenting our
numerical results for the average susceptibility and the function
$\overline{\Gamma(r)}$ along one of the principal axes. (We
believe this to be the first time, where the full three
dimensional correlation is calculated and not just parameters like
$\nu$ or $\eta$.) The results for the average susceptibility are
used to extract the critical temperature and critical exponents of
the 3D random field Ising system.
\end{abstract}
\maketitle

\section{\label{introduction}Introduction and outline}
Real space renormalization group (RSRG) served as a major tool,
over the last thirty years, in the field of critical phenomena. By
simplifying calculations near a critical point, the various RSRG
techniques, such as using the majority rule \cite{bdfn92}, the
well known Migdal-Kadanoff (MK) \cite{m75,k77}, the
Casher-Schwartz \cite{cs79} and others \cite{tmss90}, allow one to
penetrate the critical regime to a point at which critical
exponents can be extracted. For translational invariant (pure)
systems, only a single renormalization step is required to obtain
the recursion relations for the parameters of the Hamiltonian,
from which critical fixed-points may be derived along with
critical exponents. For random systems the recursion relations are
position dependent. Therefore, a natural approach is to consider
the recursion for the distribution of disorder. Equivalently,
recursion relations may be obtained for all the parameters
defining the distribution, moments, correlations, etc.
Practically, in this approach, the recursion relations are
truncated to obtain relations involving only the mean and
variance, keeping the random couplings independent
\cite{sf80,kd81,ab84,aa85}. An alternative approach, suggested
first by Berker and Ostlund \cite{bo79}, is to consider a given
realization of disorder on a finite system. Renormalization is
then used to reduce the system to a size where brute force
calculation is possible. Thermal averages of certain quantities
can thus be obtained for that realization and ensemble average is
obtained by repeating the procedure for many realizations and
averaging the results. The advantage of the method is that all the
moments and correlations generated by renormalization are kept.
The disadvantage is that the renormalization leaves in the end a
small number of spins and, therefore, only thermal averages of
functions of those spins can be evaluated directly. This is good
enough to obtain directly the ensemble average of the
magnetization \cite{dsy93}, because the average magnetization
obtained from the surviving spins is exactly the true average
magnetization. If, on the other hand, we are interested in
averages involving spins that do not survive the renormalization
process, things become much more complicated. Take for example the
ensemble averaged correlation, $\overline{\Gamma(r_{ij})} =
\overline{\langle\sigma_i\sigma_j\rangle} -
\overline{\langle\sigma_i\rangle\langle\sigma_j\rangle}$, where
$\langle\cdots\rangle$ denotes thermal average and
$\overline{\cdots}$ denotes ensemble average. It can be calculated
directly from the remaining spins provided that the vector
$r_{ij}$ connecting the sites $i$ and $j$ equals a vector
connecting two of the surviving spins or obtained from it by a
symmetry operation on the initial lattice. In any other case, a
direct calculation is impossible. It is true that quantities like
the susceptibility that involves $\Gamma_{ij}$'s for all pairs of
sites can be calculated indirectly \cite{dsy93,fbm95,fb9697,yb97}.
Following Dayan et al \cite{dsy93}, this requires adding a
constant field, $H$, to the system, calculating the ensemble
average of the magnetization as a function of $H$ by the method
outlined above and then differentiating with respect to $H$ at
$H=0$. In fact, many interesting quantities may be obtained by
adding the appropriate interaction to the Hamiltonian and
differentiating ensemble averages with respect to the
corresponding coupling constants. The trouble with that approach
is that even in the relatively simple case of evaluating the
susceptibility, the numerical differentiating is quite
problematic. Berker and coworkers \cite{fbm95,fb9697,yb97} were
using the chain rule to approximately recover thermodynamic
densities of the original system from the renormalized couplings
of the reduced system. The main problem here is that the method is
limited only to the obtainment of thermal averages of products of
spins showing in the Hamiltonian. The purpose of the present
article is to show how to calculate various interesting quantities
that involve spins that do not survive the renormalization by a
direct and effective method.

The paper is organized as follows. In Sec. \ref{SecCS}, we
describe briefly the Casher Schwartz (CS) renormalization
procedure, which is the RSRG that we use here for the numerical
demonstration of our method in Sec. \ref{SecResults}. Note,
however, that the method we present is more general and can be
used with any other renormalization scheme. In Sec.
\ref{SecMethod}, the elements of our method are mainly considered
for the random field Ising system. It is shown how to calculate
the average "connected" spin-spin correlation function,
$\overline{\Gamma_{ij}}$ and the average "disconnected" spin-spin
correlations, $\overline{S_{ij}} =
\overline{\langle\sigma_i\sigma_j\rangle}$, from which the average
susceptibility, $\overline{\chi}$, and the average total energy,
$\overline{E}$, can be easily obtained. It is further shown how to
calculate the average specific heat, $\overline{C}$, for an Ising
system with random bonds and random fields. Note that the method
presented here enables a full evaluation (though approximate) of
quantities like $\overline{\Gamma_{ij}}$ and $\overline{S_{ij}}$
that depend on distance. As we shall show later, this follows from
two facts: (a) The finite and small number of spins we are left
with at the end of the renormalization procedure. (b) The fact
that the system is random. (It will be shown how, in principle,
this method can be used to calculate by a similar method of a
finite system renormalization, such quantities in the pure system.
The practicality of the method for the pure system will prove,
however, to be questionable.) In the last section, we demonstrate
the usefulness of our method by calculating the average
susceptibility $\overline{\chi}$ and $\overline{\Gamma(r)}$ for
$r$'s lying on a main axis of the lattice, for the random field
Ising system. The evaluation of $\overline{\chi}$ is used to
derive critical exponents that may be compared with the exponents
derived by other methods. Note that the values of the exponents
depend not only on the numerical application of the method
presented here, but also on the specific scheme of renormalization
employed.

\section{\label{SecCS}Renormalization}
Although our method is general, we will use the CS scheme
\cite{cs79} for the numerical demonstration of our method and
present the results in Sec. \ref{SecResults}. Like any other
renormalization procedure (such as Migdal-Kadanoff (MK)
\cite{m75,k77} and others \cite{tmss90}), when performed on a
regular lattice, recovering the original form of the Hamiltonian
is not an exact procedure. Nevertheless, for the translational
invariant (pure) Ising system it produces good results
\cite{cs79}. In most RSRG calculations for random systems,
correlations generated by the renormalization are simply ignored.
It was suggested, however, many years ago by Harris and Lubensky
\cite{hl74} that those correlations are important. The CS scheme
generates, indeed, such correlations. Schwartz and Fishman
\cite{sf80}, used the CS scheme to renormalize the mean and
variance of the distribution of random bonds. They took into
account the generated correlations and found that inclusion of the
effect of generated correlations in the the renormalized variance
is essential.

The CS renormalization method for the pure Ising system is
described in detail in \cite{cs79}, while a detailed demonstration
of how it can be used, locally, for a random bond system can be
found in \cite{sf80}. Moreover, our numerical renormalization
procedure here, as conducted for the random field Ising model,
follows almost exactly the numerical procedure used by Dayan at al
\cite{dsy93}. Here, therefore, we only describe it in brief.
According to the CS scheme, an integration of every other site is
performed exactly, but then, all non-\textit{nn} bonds, generated
by the RG transformation, are symmetrically bent onto available
\textit{nn} bonds, many-spin odd interactions may be grouped to
form the renormalized field, while many-spin even interactions are
simply omitted. Here, though, as in Ref. \cite{dsy93}, in order to
simplify computer programming, we only keep the renormalized
fields, and ignore the many-spin odd interactions as well. The
integration of every other site is, relatively, an easy task, even
in 3D, since every spin situated on an even site interacts only
with neighboring spins situated on odd sites.

We start then with a set of $N=L^3$ Ising spins, $\sigma_i=\pm 1$,
with $L=2^n$, situated on a 3D SC lattice. Suppose, now, that the
Ising system is not translational invariant and represented by the
Hamiltonian
\begin{equation}
{\cal H}=-\sum_{<i,j>}J_{ij}\sigma_i\sigma_j-\sum_i h_i\sigma_i,
\label{EqFieldHam}
\end{equation}
where $<i,j>$ refers to nearest neighbors only. Performing the
trace over every other site, each of the erased spins, contributes
separately to each of the terms in the new Hamiltonian, generating
all possible interactions among its 6 \textit{nn}'s. We arrive,
then, at a new Hamiltonian, containing fields, \textit{nn},
\textit{nnn} and multi-spin interaction terms, from 3-spin and up
to 6-spin interactions. All the couplings are again local. The
result is even further complicated by the fact that the resulting
lattice is not an SC but an FCC lattice. As was mentioned above,
the multi-spin interactions are simply ignored, while the values
of the 3 generated \textit{nnn} interactions are symmetrically
distributed over the 12 \textit{nn} interactions. To bring the
lattice back to its SC form, we still need to integrate over each
of the face-centered spins. To do that, we first bend all
\textit{nn} bonds connecting between face-centers, onto
\textit{nn} bonds that lie on the face of the cube and connecting
between face-centers and vertices. The extra erasure step can now
be easily executed following the CS 2D renormalization scheme.
This is a much simpler procedure, which we shell not describe
here, but can also be found in Refs. \cite{cs79,sf80}.

In our study, the above two steps procedure is performed locally
and repeated iteratively until the system is brought down to a
size of $2\times 2\times 2$, for which a trace can be performed
exactly.

\section{\label{SecMethod}The method}
\begin{figure}[h]
\includegraphics{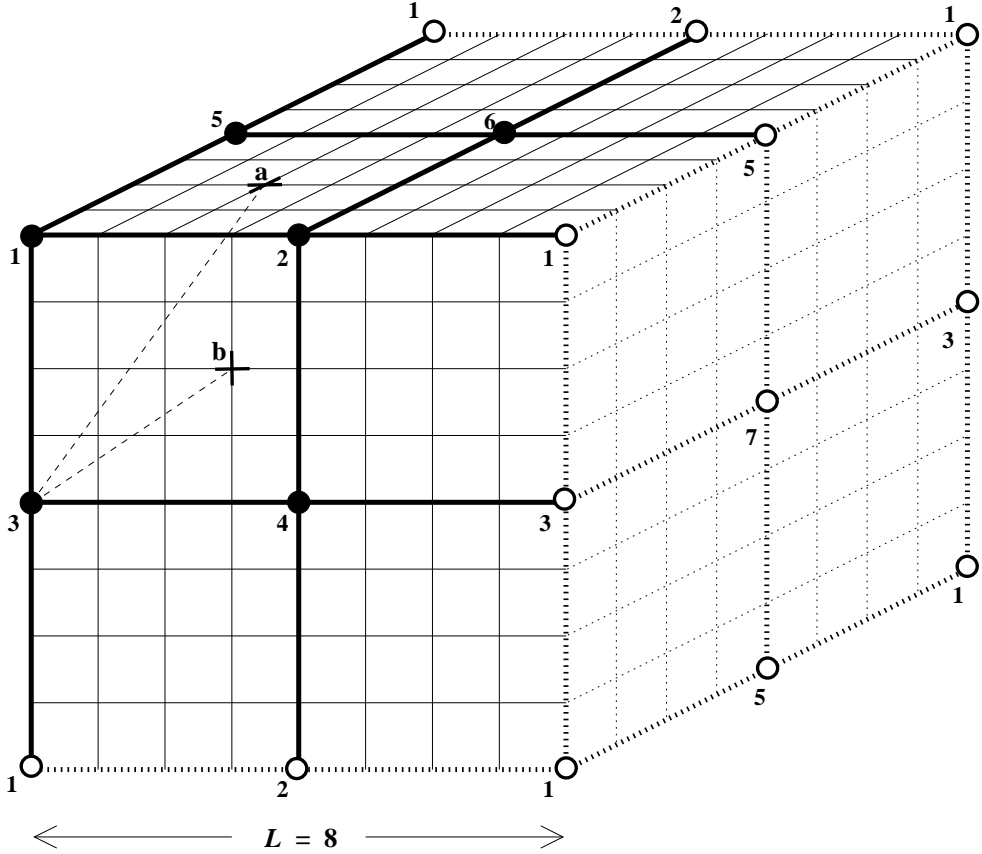}
\caption{\label{FigSC} Starting with a large system of linear size
$L=2^n$ (here demonstrated with $n=3$), the system is fully
reduced, using some RSRG transformation, to its minimal linear
size of $L=2$. The black numbered dots and the thickened lines
connecting them, indicate, respectively, the 8 remaining sites
(only 6 of them are shown) and the remaining bonds connecting
them, of the renormalized system. The circles and the doted lines
reflect the boundary conditions imposed on the system. Sites that
are \textit{nn}'s in the reduced system (such as 1 and 2) are, in
fact, $L/2$ lattice constants apart in the original system. From
all distances available in the original system (such as
$3\leftrightarrow a$ and $3\leftrightarrow b$), only $L/2$ for
\textit{nn}'s, $L/\sqrt{2}$ for \textit{nnn}'s (like
$1\leftrightarrow 4$) and $\sqrt{3}L/2$ on the main diagonals
(like $3\leftrightarrow 6$) are available in the reduced system.}
\end{figure}

In this section we describe, mainly, how to calculate the
"connected" and "disconnected" correlation function for the three
dimensional random field system. We consider the random field
Hamiltonian
\begin{equation}
{\cal H}=-J\sum_{<i,j>}\sigma_i\sigma_j-\sum_i h_i\sigma_i.
\label{EqFieldHamJ}
\end{equation}
The $h_i$'s are random uncorrelated fields, distributed around
zero,
\begin{equation}
\overline{h_i}=0, \qquad \overline{h_i h_j} = h^2\delta_{ij},
\label{EqMeanStndrdDev}
\end{equation}
We assume that the random fields are distributed according to a
Gaussian distribution,
\begin{equation}
P\{h\} = \prod_i P_i (h_i) \equiv
\frac{1}{\Lambda}e^{-\frac{1}{2h^2}\sum_i{h_i}^2},
\label{EqGaussDist}
\end{equation}
where $\Lambda \equiv (h \sqrt{2\pi})^N$. We are interested in
calculating the average spin-spin correlations and susceptibility
of a large, but finite, system, over a large number of
realizations of the random field. Consider first
\begin{equation}
\Gamma_{ij} \equiv \langle\sigma_i\sigma_j\rangle -
\langle\sigma_i\rangle\langle\sigma_j\rangle.
\label{EqGammaij}
\end{equation}
The susceptibility is related to the spin-spin correlations by
\begin{equation}
\chi=\frac{\beta}{N}\sum_{i,j}\Gamma_{ij}.
\label{EqSusc}
\end{equation}
The average susceptibility,
\begin{equation}
\overline{\chi}=\beta\sum_j\overline{\Gamma_{ij}},
\label{EqAvSusc}
\end{equation}
is obtained by averaging over a large enough number of
realizations. The true average of $\Gamma_{ij}$ depends, of
coarse, only on the radius separating $i$ and $j$. Namely,
translational invariance is restored by averaging. Obviously,
since we consider a large system, calculating the above quantities
directly involves the impossible task of performing a trace over a
large number of spins numerically. We, thus, turn to real space
renormalization. By choosing first a specific renormalization
scheme (CS, MK, etc.), the rescaling factor, $b$, is set. We then
choose the linear size of our system, $L$, to be $b$ to some
integer power, $n\geq 2$, depending on how large we want it to be.
The renormalization transformation is then used locally, by
performing $n-1$ repeated iterations, to fully reduce the size of
the system to $b\times b\times b$. The thermal average of each of
the 8 remaining spins can now be calculated exactly by performing
the trace using the Hamiltonian of the reduced system. In fact,
averaging over large enough number of realizations, we may expect
$\overline{\langle\sigma_i\rangle}$ to be translational invariant
and, therefore, to be equal to the average magnetization per spin.
In practice, in order to improve our averaging, we will calculate
it as follows,
\begin{equation}
\overline{M}\equiv
\frac{1}{N}\sum_{i=1}^N\overline{{\langle\sigma_i\rangle}}_{{\cal
H}_N} =
\frac{1}{8}\sum_{i=1}^8\overline{{\langle\sigma_i\rangle}}_{{\cal
H}_8}.
\label{EqM}
\end{equation}
Although, in principle, calculating thermal averages for the
renormalized  8-spin system is a reasonable task, we are now faced
with a different problem. The problem is that the reduced system
only carries information about the 8 spins that survived the
renormalization procedure. While this makes no difference when
calculating quantities containing thermal averages of a single
spin (such as the average magnetization above), it makes it
impossible to calculate directly quantities that contain thermal
averages of more than one spin. Such are the average spin-spin
correlations, $\overline{\Gamma_{ij}}$, at distances other than
$L/2$, $L/\sqrt{2}$ and $\sqrt{3}L/2$ and such is the average
susceptibility, which, according to Eq. (\ref{EqAvSusc}), requires
the sum of $\Gamma_{ij}$ over all distances available in the
original system. This is simply because spins that are
\textit{nn}'s in the reduced system are, in fact, $L/2$ lattice
constants apart in the original system (see Fig. \ref{FigSC}).
Indeed, as was done by Dayan at al \cite{dsy93}, one may calculate
the susceptibility by applying a small external field, $H$, to the
original system and then use the derivative of $\overline{M}$ with
respect to $H$. This method, though, is quite problematic, because
it concerns a numerical derivative at $H=0$. As was discussed by
Dayan at al, $H$ must be small enough, so that $M(H)$ is linear in
$H$. This is difficult to achieve, since, below the transition,
the size of the region where that linearity exists, shrinks to
zero as the size of the system tends to infinity. If, on the other
hand, the field is too small, one may encounter numerical problems
from round-off errors.

Our solution to the problem is obtained by using the identity:
\begin{equation}
\overline{\Gamma_{ij}}= \frac{1}{\beta
h^2}\overline{\langle\sigma_i\rangle h_j}.
\label{EqTheorem1SS}
\end{equation}
This identity was used in the past
\cite{ss85,ss86,sgn91,gaahs9396} but since its proof is very short
and simple, we will derive it here again for the sake of
completeness of the presentation. Start with the right hand side
of Eq. (\ref{EqTheorem1SS}).
\begin{eqnarray}
\frac{1}{\beta h^2}\overline{\langle\sigma_i\rangle h_j} &=& -
\frac{1}{\beta}\int\langle\sigma_i\rangle\frac{\partial}{\partial
h_j} P\{h\} D h \nonumber \\
&=& \frac{1}{\beta}\int\frac{\partial
\langle\sigma_i\rangle}{\partial h_j} P\{h\} D h.
\label{EqProof}
\end{eqnarray}
This completes the proof since it is easy to see, using the random
field Hamiltonian \ref{EqFieldHamJ}, that
\begin{equation}
\frac{1}{\beta}\frac{\partial \langle\sigma_i\rangle}{\partial
h_j}=\langle\sigma_i \sigma_j\rangle - \langle\sigma_i \rangle
\langle\sigma_j \rangle.
\label{EqDerivEqCorr}
\end{equation}

\subheading{Connected spin-spin correlations and susceptibility}
To obtain $\overline{\Gamma(\overrightarrow{r})}$, the following
procedure is used. The thermal average of $\sigma_i$ is calculated
in a given realization for $i$ that is one of the surviving spins.
it is, then, multiplied by the value of $h_j$ in that realization,
where $j$ is a site separated by a vector $\overrightarrow{r}$
from $i$ on the original lattice. The product $\langle\sigma_i
\rangle h_j$ is then averaged over many realizations. Since the
true average should depend only on the vector connecting the sites
and that up to a symmetry of the lattice, the statistics can be
considerably improved by averaging
$\overline{\Gamma(\overrightarrow{r})}$ as follows,
\begin{equation}
\overline{\Gamma(\overrightarrow{r})}= \frac{1}{\beta
h^2}\frac{1}{8}\overline{\sum_{i=1}^8\frac{1}{n_r}\langle\sigma_i
\rangle\sum_{k=1}^{n_r} h_{i_k}}, \label{EqTheorem1SSNum}
\end{equation}
where $i_k$, runs over all equivalent sites around $i$, that are
at a distance $r$ from it. The number of these sites is $n_r$. In
our numerical study, presented in Sec. \ref{SecResults}, we have
limited ourselves to correlations in the directions of the
principal axes of the lattice, so that $n_r=6$ (except, of coarse,
for self correlations, where $r=0$ and $n_r=1$). For the average
susceptibility, the statistics is improved by writing
\begin{equation}
\overline{\chi}=\frac{1}{8h^2}\overline{\sum_{i=1}^8
\langle\sigma_i\rangle\sum_j h_j}.
\label{EqAvSuscNum}
\end{equation}

In the following we will show how to evaluate "disconnected"
correlation functions. Since, in practice, this involves much
heavier computations, we just describe how it should be done, and
postpone actual numerical application to future publications.

\subheading{The 'sites translation' method for the "disconnected"
spin-spin correlations and the total average energy} The main
point in the evaluation of "disconnected" correlations is that,
actually, the method of integrating out many degrees of freedom
and remaining with a small number of spins can yield not only
$\sigma_i$, where $i$ is a surviving spin, but indeed all the
local magnetizations. Namely, the method enables to calculate
$\sigma_i$ for all $i$ in the original lattice. This can be done
by noting that, for a given realization, we can translate the 8
surviving spins. (This can be done, equivalently, by keeping the
surviving spins and translating the field configuration.)
Therefore, we choose a realization, evaluate the 8
$\langle\sigma_i\rangle$'s corresponding to the surviving spins,
then translate the field configuration by one lattice spacing,
thus obtaining the local magnetization of the 8 sites translated
from the original set by one lattice spacing in the opposite
direction. This is repeated until all the local magnetizations are
obtained. As is easily seen, this involves order of $N$
repetitions of the original procedure. It is clear now how, by
obtaining all the $\langle\sigma_i\rangle$'s, we can obtain
\begin{equation}
\overline{S_{ij}} = \overline{\Gamma_{ij}}+
\overline{\langle\sigma_i\rangle\langle\sigma_j\rangle}=
\frac{1}{\beta h^2}\overline{\langle\sigma_i\rangle h_j} +
\overline{\langle\sigma_i\rangle\langle\sigma_j\rangle}.
\label{EqSSDisConCorr}
\end{equation}
This is generally time consuming but not so bad if we are
interested, say, in the energy given by
\begin{equation}
\overline{E} =
-J\sum_{<i,j>}\overline{\langle\sigma_i\sigma_j\rangle}-\sum_i
\overline{ h_i\langle\sigma_i\rangle}, \label{EqTotEnergy}
\end{equation}
for which our method yields the following expression
\begin{equation}
\overline{E} = -J\sum_{<i,j>}\overline{\left(\frac{h_j}{\beta h^2}
+ \langle\sigma_j\rangle\right) \langle\sigma_i\rangle}-\sum_i
\overline{ h_i\langle\sigma_i\rangle}.
\label{EqTotEnergyEx}
\end{equation}
The calculation involves taking the ensemble average of the
magnetization multiplied by the field at the same point (which is
just $\overline{\Gamma_{ii}}$) and the product
$\langle\sigma_i\rangle\langle\sigma_j\rangle$, where $i$ and $j$
are nearest neighbors. To obtain the same degree of accuracy in
$\overline{S_{ij}}$ as in $\overline{\Gamma_{ij}}$, we need for
each field configuration, to perform seven renormalization
procedures instead of the one needed for calculating
$\overline{\Gamma_{ij}}$. The reason is that we need the original
set of surviving spins and the sets obtained from it by the six
unit translations in all directions. For other "disconnected"
correlation, the time factor needed to attain the accuracy of the
"connected" correlation is $n_r+1$.

\subheading{Specific heat for systems of random bonds and fields}
Next, we show how to obtain the average specific heat for an Ising
Hamiltonian with both, random fields and random bonds. The
evaluation of the average specific heat,
$\overline{C}=kT^2\left(\overline{\langle{\cal
H}^2\rangle}-\overline{\langle{\cal H}\rangle^2}\right)$, is also
made possible by our method. It is given by
\begin{equation}
\overline{C}=\frac{N}{k\beta^2}\left(\frac{N}{2}\sum_{<k,l>}
\overline{J_{ij}J_{kl}\Gamma_{(ij)(kl)}} + 2\sum_{<k,l>}
\overline{h_i J_{kl}\Gamma_{(i)(kl)}} + \sum_j\overline{h_i
h_j\Gamma_{ij}}\right),
\label{EqSpcHeat}
\end{equation}
where for $A$ and $B$, two sets of indices, we define $\Gamma_{AB}
\equiv \langle\sigma_A\sigma_B\rangle -
\langle\sigma_A\rangle\langle\sigma_B\rangle$ and
$\sigma_A\equiv\prod_{i \in A}\sigma_i$. The parenthesis in the
subscripts of $\Gamma$ are used to describe the sets of spins.
Operating on the three terms of Eq. (\ref{EqSpcHeat}) (from the
less complicated on the right to the more complicated on the
left), basically using integration by parts, we obtain first,
\begin{subequations}
\label{EqSpecHeatTerms}
\begin{eqnarray}
\overline{h_i h_j\Gamma_{ij}} &=&
\overline{h_i\left(h_j\Gamma_{ij}+\frac{1}{\beta}\langle\sigma_i
\rangle\right) - \frac{1}{\beta}h_i\langle\sigma_i \rangle} \nonumber\\
&=&\frac{1}{\beta}\overline{h_i\left[\frac{\partial}{\partial h_j}
\left(h_j\langle\sigma_i \rangle\right)-\langle\sigma_i
\rangle\right]} =
\frac{1}{\beta}\overline{\left(\frac{{h_j}^2}{h^2}-1\right)
h_i\langle\sigma_i \rangle}.
\label{EqSpecHeatTerm3}
\end{eqnarray}
The next two terms are obtained similarly by following the same
steps. The only difference is that, since we assume that the
random bonds are distributed around some mean value, $\mu_J$, and
uncorrelated with a standard deviation $\Delta_J$, as a
preliminary step, $J_{ij}$ and $J_{kl}$ are replaced with
$\mu_J+\delta J_{ij}$ and $\mu_J+\delta J_{kl}$ respectively.
These are collected back by the end of the calculation. The
results are similar to Eq. (\ref{EqSpecHeatTerm3}) only with the
appropriate indices,
\begin{equation}
\overline{h_i J_{kl}\Gamma_{(i)(kl)}} =
\frac{1}{\beta}\overline{\left(\frac{J_{kl}\delta
J_{kl}}{{\Delta_J}^2}-1\right)h_i\langle\sigma_i \rangle}
\label{EqSpecHeatTerm2}
\end{equation}
and
\begin{equation} \overline{J_{ij}J_{kl}\Gamma_{(ij)(kl)}} =
\frac{1}{\beta}\overline{\left(\frac{J_{kl}\delta
J_{kl}}{{\Delta_J}^2}-1-\delta_{ij,kl}\right)J_{ij}\langle\sigma_i
\sigma_j\rangle}. \label{EqSpecHeatTerm1}
\end{equation}
\end{subequations}
Now, Eq. (\ref{EqSpecHeatTerm1}) is not yet in its final form,
since it still contains the term $\langle\sigma_i\sigma_j\rangle$.
We thus first need to use Eq. (\ref{EqSSDisConCorr}) in order to
fix that, and then Substitute Eqs. (\ref{EqSpecHeatTerms}) back
into Eq. (\ref{EqSpcHeat}).

It may seem plausible that techniques of the nature described
above can be used also for correlations of quantities coupled by
position independent coupling constants (that may be even zero).
In principle, this is true because we can always add random
couplings with a Gaussian distribution, do the calculation and, in
the end, take the variance of those couplings to zero.
practically, it is unclear whether such a procedure is more
effective than taking a numerical derivative, since both imply
taking the limit where certain couplings tend to zero. (In the
random field problem, with no random bonds, we can either add
random bonds with variance that will eventually tend to zero, or
take a numerical derivative with respect to $T$.)

\section{\label{SecResults}Numerical results}
To demonstrate the usefulness of our method, we present in this
section results for the less time consuming quantities. we present
calculation of the average susceptibility and derive from it the
critical exponents, comparing the results with those of Ref.
\cite{dsy93}, which uses the same renormalization scheme. We also
present, for the first time we believe, a calculation of an
$\overrightarrow{r}$ dependent correlation. We evaluate
$\Gamma(\overrightarrow{r})$ as a function of temperature and
distance, for $\overrightarrow{r}$'s on the principal axes of the
lattice. All figures are presented with error bars, although, in
some of them, the error bars are too small to be noticed. The
errors are the standard deviation calculated from the data.

\begin{figure}[p]
\includegraphics[width=.88\textwidth]{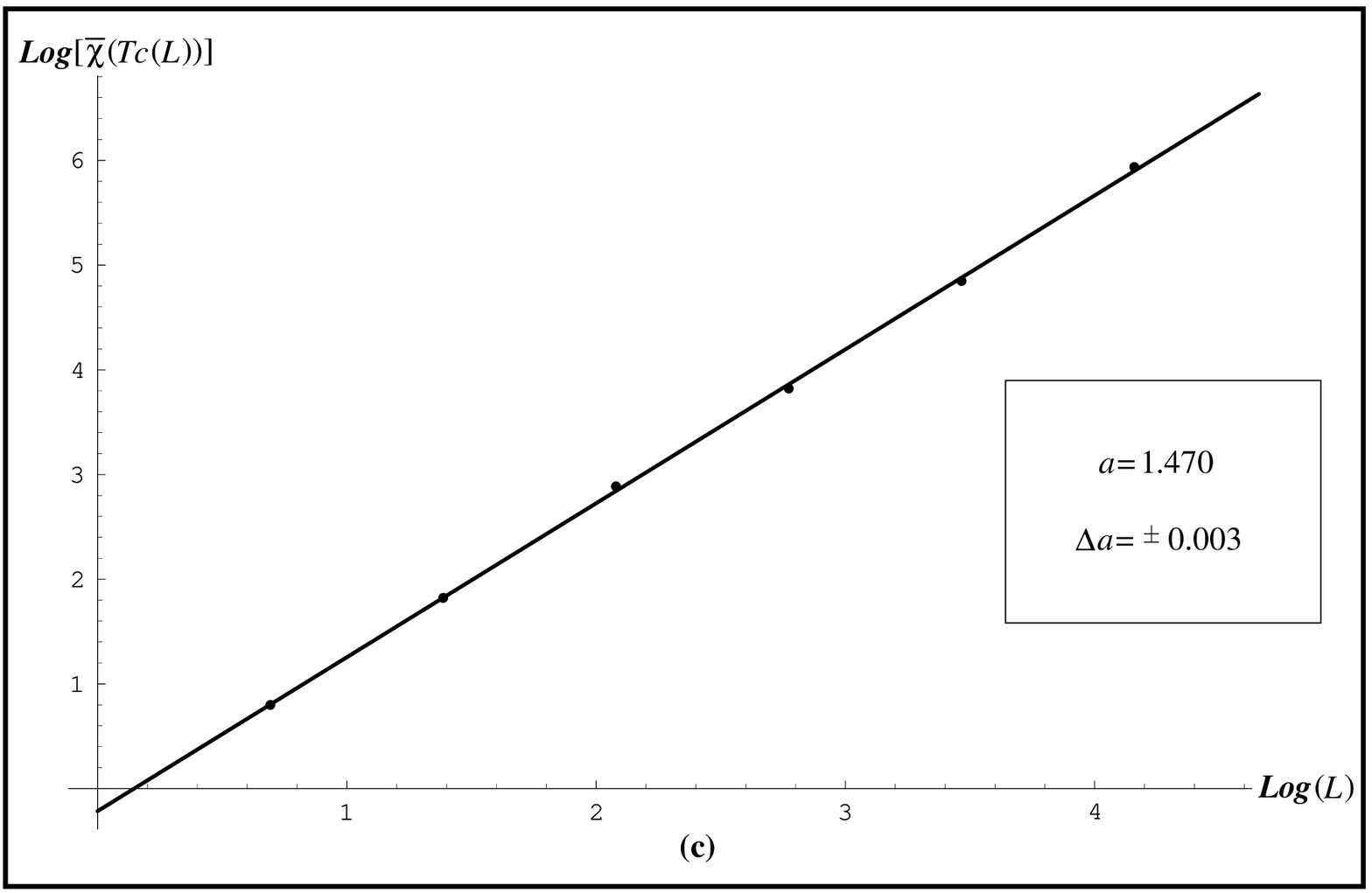}
\caption{\label{FigSusch1} In (a), the average susceptibility,
$\chi$, is shown as a function of temperature, $T$, measured in
units of $1/\beta$. It is presented in a $Log$ plot. The external
field, $H$, is zero while the standard deviation of the random
field is $h=1$. As indicated on each figure, the different levels
of grayscaling correspond to systems of different linear size,
$L$. While for $L$'s up to $64$, the system is averaged over
$10000$ realizations, the largest, $L=128$, system is averaged
only over $1000$ realizations. This may be the cause for the
upwards displacement of the average susceptibility for the
largest, $L=128$, system. In (b), excluding the largest system,
the logarithm of the maximum of the susceptibility is plotted
verses the logarithm of $L$. The value of $\eta$ is obtained from
a linear fit, as, by Eq. (\ref{EqSuscFinSizScal}), the slope $a$
is $\gamma/\nu=2-\eta$.}
\end{figure}

Using the CS approximation, we choose our input parameters, for
calculating the average susceptibility, equal to those used by
Dayan at al \cite{dsy93}, who also used the CS scheme. We also
partially follow their line of analysis for extracting the
critical exponents $\eta$, $\gamma$ and $\nu$. This serves as a
baseline, with which part of our results may be compered. We,
thus, set $J=1$ and $h=1$ in the Hamiltonian (\ref{EqFieldHamJ})
and use $1/\beta$ as temperature. We have calculated the
susceptibility as a function of temperature for systems of linear
sizes $L=2,4,8,16,32,64$ and $128$, averaging, each, over 10000
realizations of the random field, except for the largest system,
for which we had to be satisfied with only 1000 realizations. This
is shown in Fig. \ref{FigSusch1}(a) in a $Log$-plot. Like Dayan at
al, we also have included the mirror image of each random field
realization, and that is in order to preserve, in our finite
systems, the basic reflection symmetry of the infinite system. As
seen in the figure, there is an upwards displacement of the
average susceptibility for the largest, $L=128$, system. This is
probably because the relatively small number of $N=1000$
realizations, for that system, is too little to statistically rely
on. It may also be that, since the renormalization procedure in an
approximation, the large number of renormalization steps generates
an error which is too large. It is, therefore, excluded in the
following analysis concerning the susceptibility. From Fig.
\ref{FigSusch1}(b), we extract the value of $\eta$, using the
finite size scaling behavior of the susceptibility,
\begin{equation}
\chi(T_{c}(L)) \sim L^{\gamma/\nu},
\label{EqSuscFinSizScal}
\end{equation}
together with the scaling relation
\begin{equation}
\gamma=(2-\eta)\nu.
\label{EqScalRelat}
\end{equation}
We thus obtain $\eta=0.53 \pm 0.003$, where the error is the
statistical one and errors generated by the renormalization
approximate procedure are, therefore, not taken into account.

\begin{figure}[p]
\includegraphics[width=.88\textwidth]{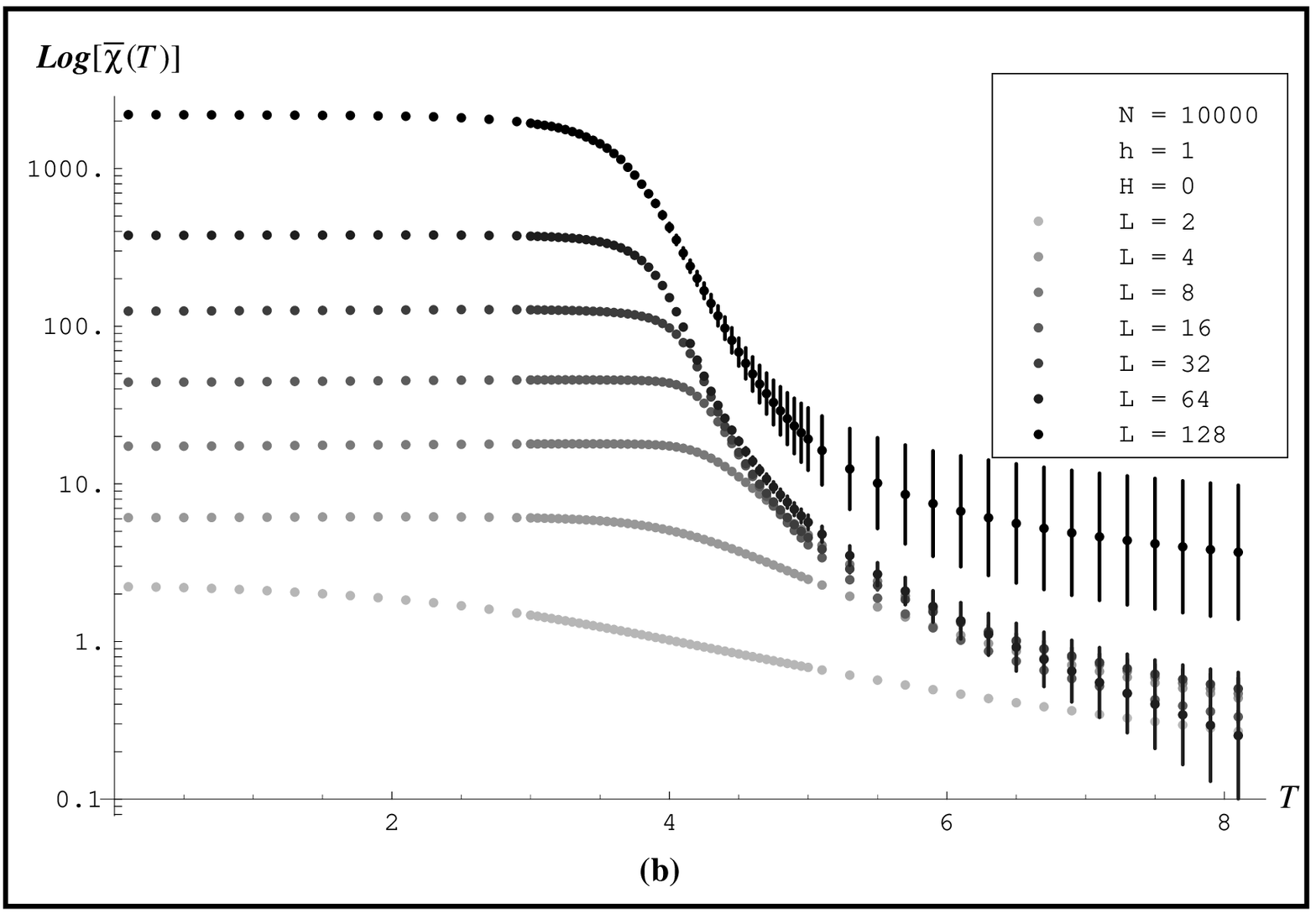}
\end{figure}

\begin{figure}[p]
\includegraphics[width=.88\textwidth]{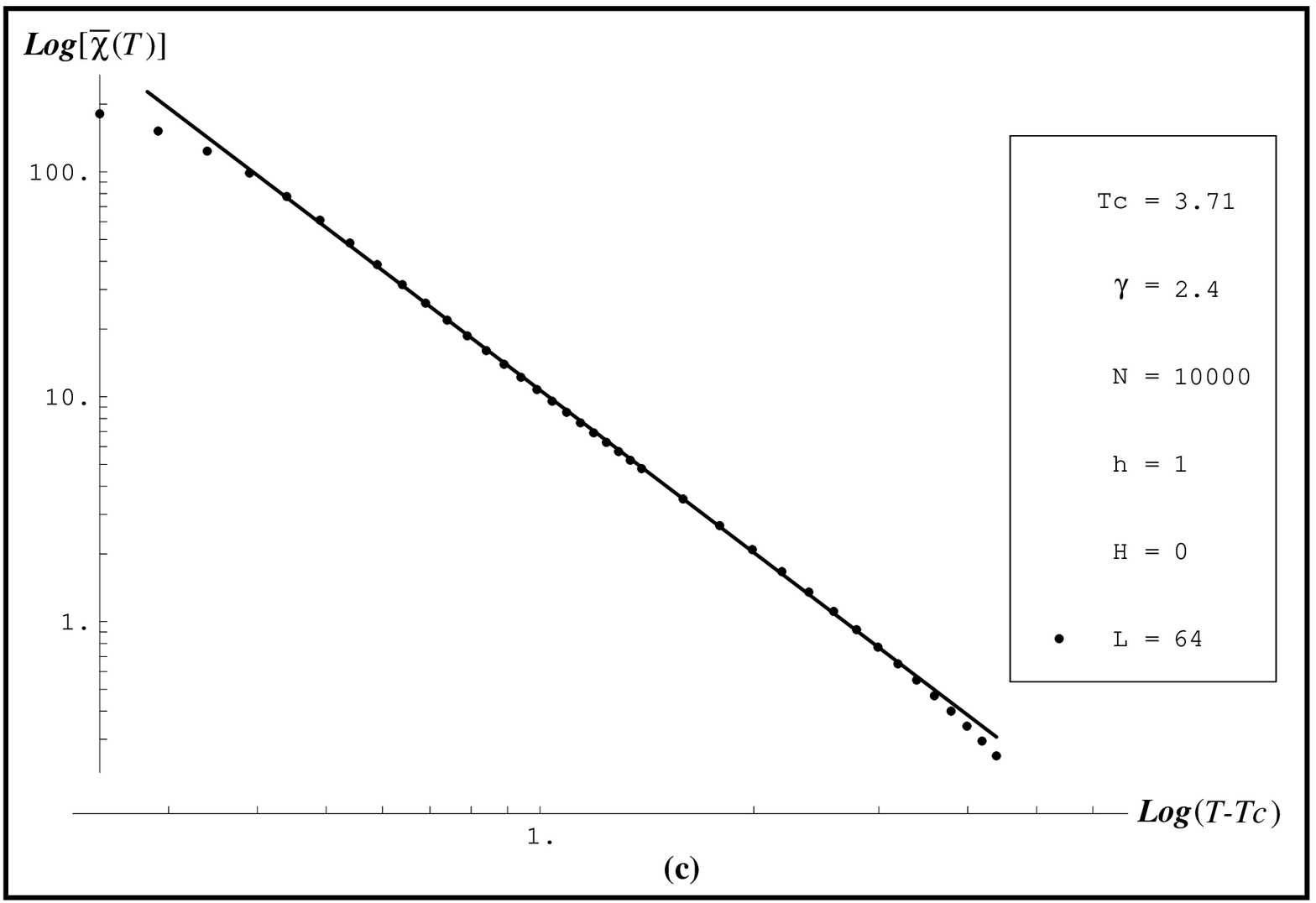}
\caption{\label{FigSuscFitGamah1} The logarithm of the average
susceptibility of a large system ($L=64$) is shown as a function
of the logarithm of $T-T_c$, for temperatures above $T_{c}$ and
for three different values of $T_c$. The two extreme, but yet
acceptable, values of $T_c$, as discussed in the text, are
presented in (a) and (c). It appears that the largest linear
region is obtained for $T_c=3.8$, as shown in (b). In (c), one may
already notice that linearity breaks by the appearance of an
opposite curvature. The value of $\gamma$ for a given $T_c$ is
determined by taking the logarithm of both sides of Eq.
(\ref{EqSuscCritBehav}).}
\end{figure}

To estimate the critical temperature, $T_c$, of the infinite
system together with the critical exponent $\gamma$,  The
logarithm of the average susceptibility of the largest system
($L=64$) is plotted verses the logarithm of $T-T_c$, for
temperatures above $T_c$ and for different values of $T_c$.
Acceptable values of $T_c$ are such that, by lowering the
temperature towards $T_c$, the graph enters a linear region until
finite size effects become important. We find that $3.71\leq
T_{c}\leq 3.95$. The two extremes are presented in Figs.
\ref{FigSuscFitGamah1}(a) and \ref{FigSuscFitGamah1}(c). For
critical temperatures above $T_c=3.95$ the linear region
disappears, while for critical temperatures around $3.71$, an
opposite curvature begins to appear, as demonstrated by Fig.
\ref{FigSuscFitGamah1}(c) for $T_c=3.71$. We consider $T_c=3.8$
[Fig. \ref{FigSuscFitGamah1}(b)], for which the largest linear
region is obtained, to be the more probable value for the critical
temperature. Next, the value of $\gamma$ for a given $T_c$ is
estimated using the critical behavior of the susceptibility,
\begin{equation}
\chi(T) \approx A|T-T_c|^{-\gamma},
\label{EqSuscCritBehav}
\end{equation}
where $A$ is some constant. Taking the logarithm of both sides of
Eq. (\ref{EqSuscCritBehav}), for a given $T_{c}$, the resulted
straight line is, then, fitted to the linear region by varying
$\gamma$ and the constant $A$. As indicated by Fig.
\ref{FigSuscFitGamah1}, the range of acceptable $T_c$'s
corresponds to a range of possible values for $\gamma$: $1.9\leq
\gamma\leq 2.4$, which is quite similar to that obtained by Dayan
at al \cite{dsy93}. Using the scaling relation
(\ref{EqScalRelat}), the resulting values for the critical
exponent, $\nu$, are roughly: $1.3\leq \nu\leq 1.65$. For
$T_c=3.8$, the corresponding values for $\gamma$ and $\nu$ are
$\gamma=2.2$ and $\nu=1.5$.
\begin{figure}[p]
\includegraphics[width=.88\textwidth]{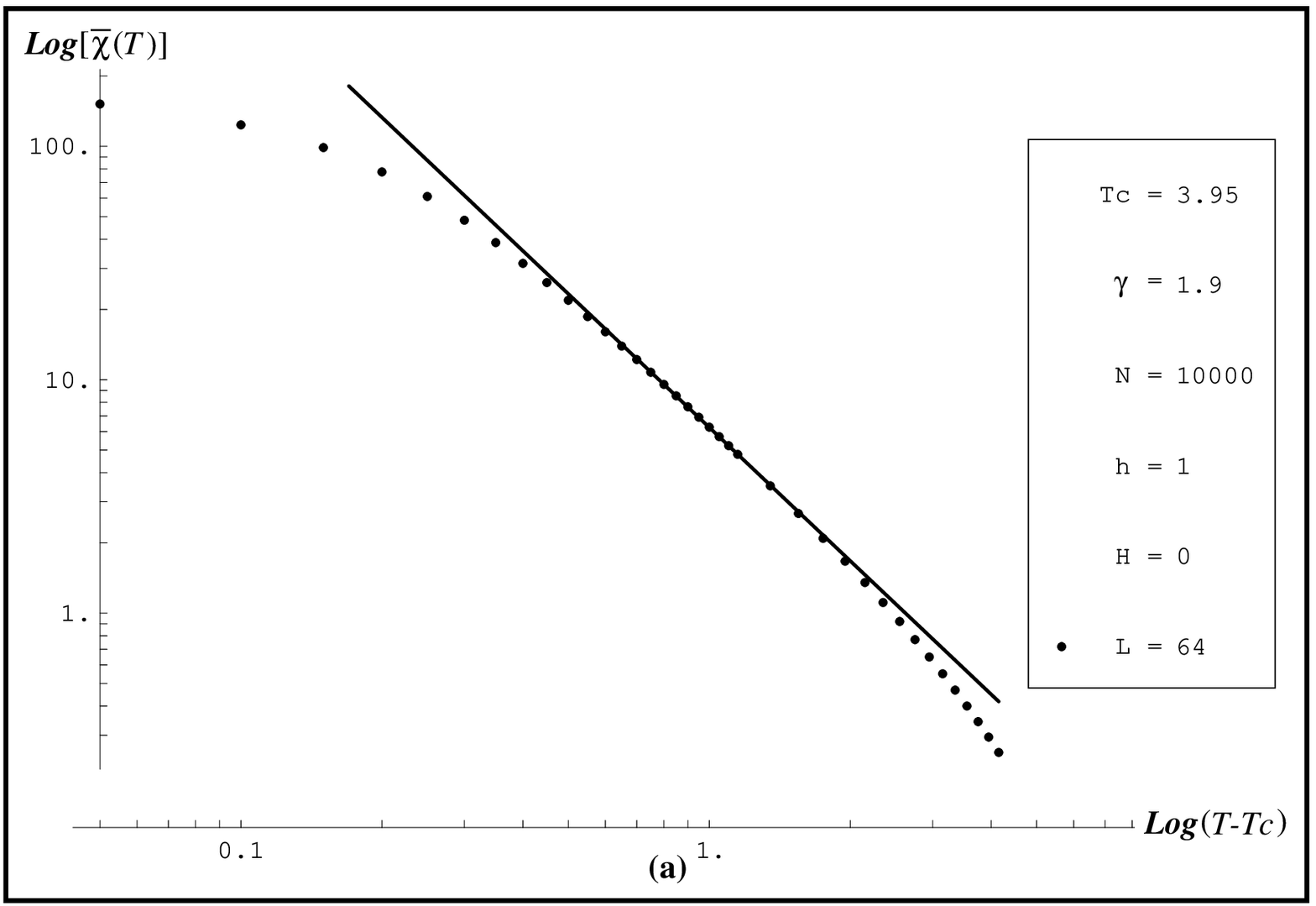}
\end{figure}
\begin{figure}[p]
\includegraphics[width=.88\textwidth]{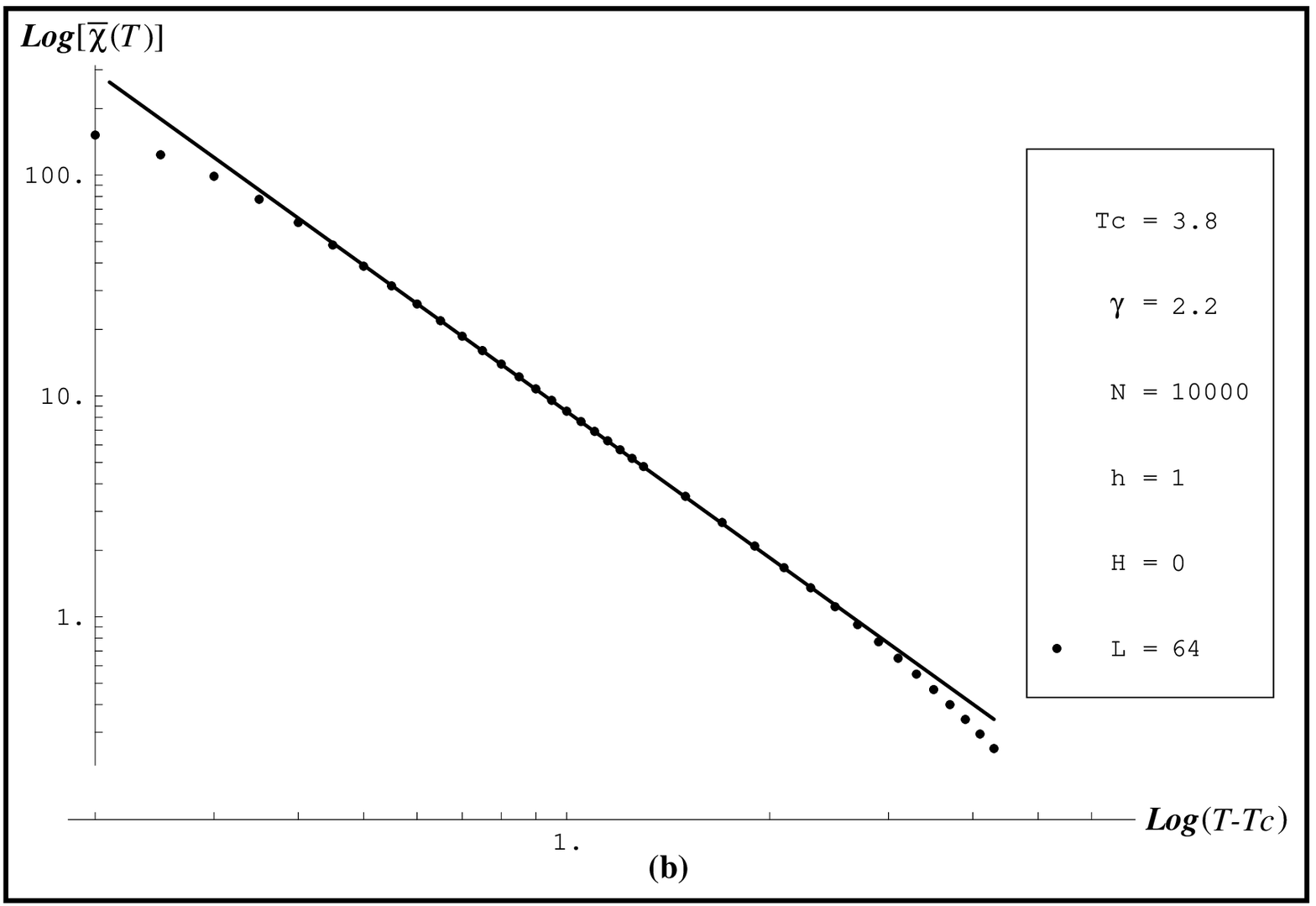}
\end{figure}

\begin{figure}[p]
\includegraphics[width=.88\textwidth]{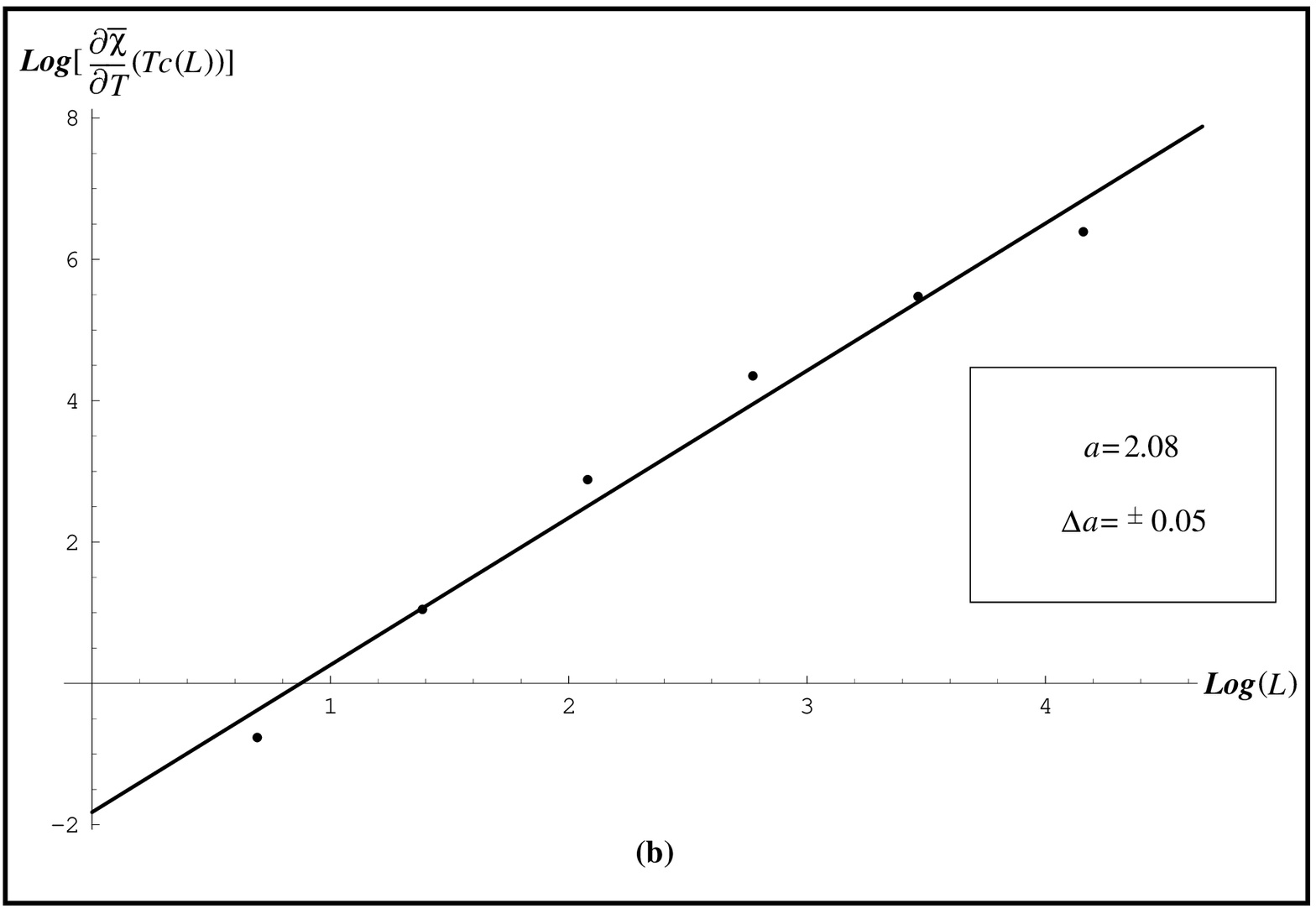}
\caption{\label{FigDSuscDTh1} In (a), the average of the absolute
value of the derivative of the susceptibility is shown as a
function of temperature. The external field, $H$, is zero while
the standard deviation of the random field is $h=1$. As indicated
on each figure, the different levels of grayscaling correspond to
systems of different linear size, $L$. In (b), the logarithm of
the maximum of $-\overline{\partial\chi/\partial T}$ is plotted
verses the logarithm of $L$. The value of $(\gamma+1)/\nu$
obtained from the slope of the linear fit here, together with
$\gamma/\nu$ obtained from the linear fit presented in Fig.
\ref{FigSusch1}(b), determine the values for $\gamma$ and $\nu$.}
\end{figure}

In a search for a more refined estimation of gamma, we turn to the
absolute value of the derivative of $\chi$ with respect to $T$.
Here we use the numerical derivative of our data for the average
susceptibility, although one should note that it is, probably,
better calculated directly from
\begin{equation}
\overline{\frac{\partial\chi}{\partial T}}=
\frac{k\beta^2}{h^2}\overline{[\langle{\cal H}\sigma_i\rangle -
\langle{\cal H}\rangle\langle\sigma_i\rangle] \sum_j h_j},
\label{EqAvDSuscDT}
\end{equation}
following our method and using the ideas presented in Sec.
\ref{SecMethod}. Similar to the finite size scaling behavior of
$\chi$, as given by Eq. (\ref{EqSuscFinSizScal}), its derivative
with respect to $T$ is expected to behave as
\begin{equation}
\frac{\partial\chi}{\partial T}(T_c(L)) \sim L^{(\gamma+1)/\nu}.
\label{EqDSuscDTFinSizScal}
\end{equation}
Here $T_c(L)$ is the temperature of the maximum of the derivative
for a given $L$ and is, therefore, different from that of the
maximum of $\chi$, used in Eq. (\ref{EqSuscFinSizScal}). We have
calculated, then, $-\overline{\partial\chi/\partial T}$ as a
function of temperature for systems of linear sizes
$L=2,4,8,16,32,64$, as presented in Fig. \ref{FigDSuscDTh1}(a). In
Fig. \ref{FigDSuscDTh1}(b), we have used a linear fit for the
logarithm of the maximum of $-\overline{\partial\chi/\partial T}$
plotted versus the logarithm of $L$, from which we obtain the
value of $(\gamma+1)/\nu$. That, together with $\gamma/\nu$
obtained earlier from the finite size scaling of $\chi$ [Fig.
\ref{FigSusch1}(b)], determine the values of $\gamma$ and $\nu$.
we obtain $\gamma=2.41\pm 0.07$ and $\nu=1.64\pm 0.07$. Note that
together with our previously obtained value of $\eta=0.53\pm
0.003$, the scaling relation (\ref{EqScalRelat}) is satisfied.

\begin{figure}[p]
\includegraphics[width=.88\textwidth]{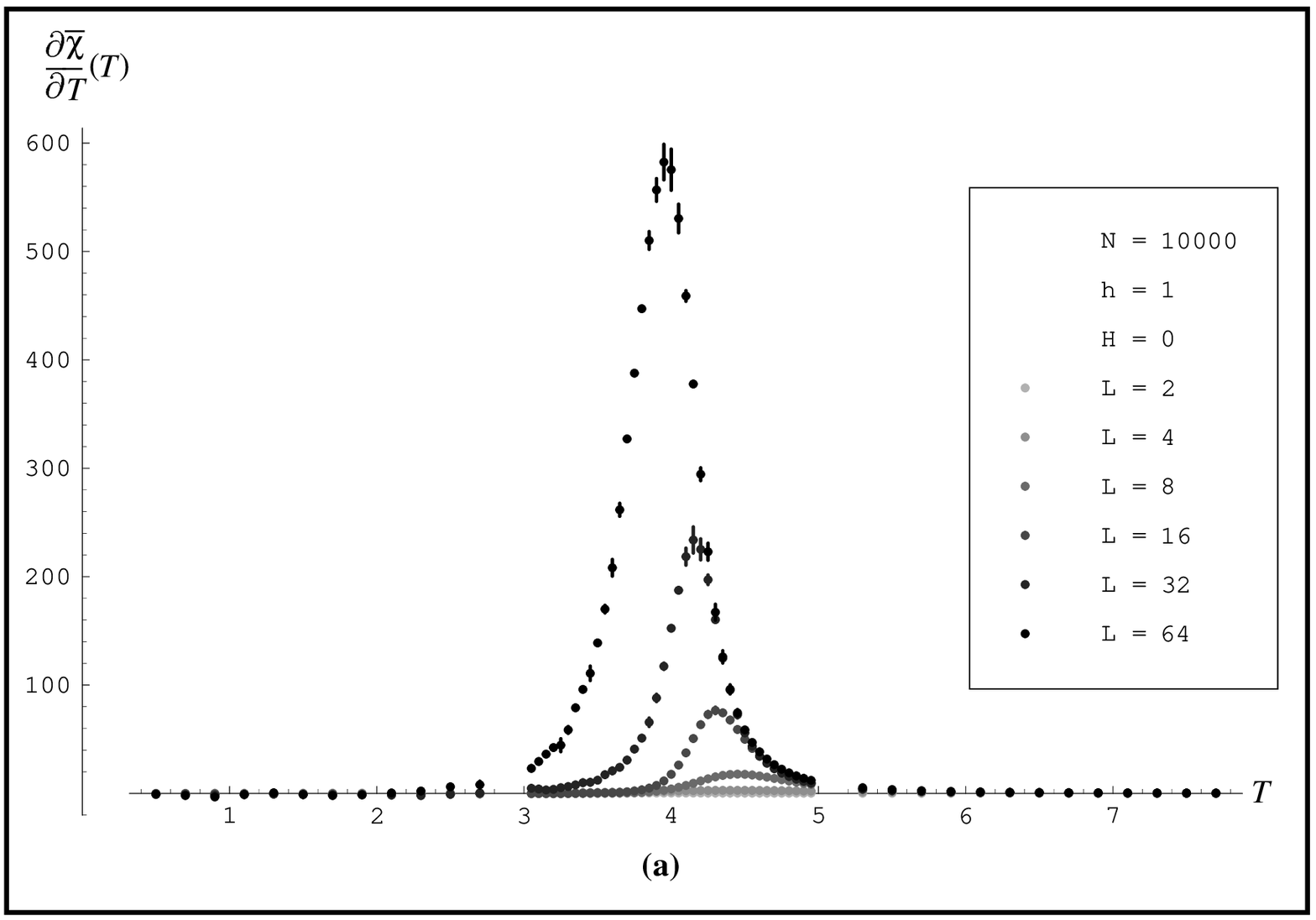}
\end{figure}

\begin{figure}[p]
\includegraphics[width=.88\textwidth]{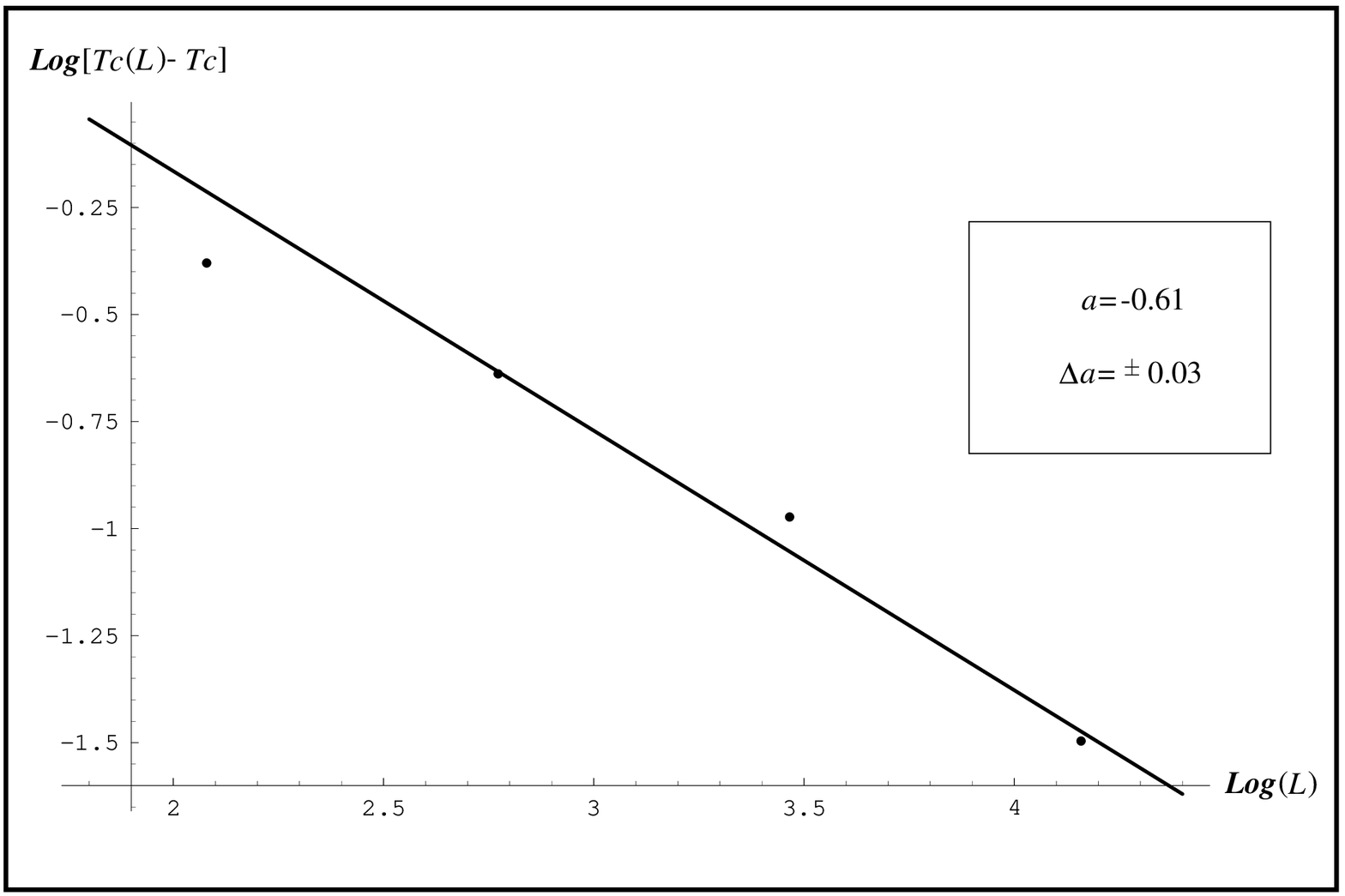}
\caption{\label{FigFitTcL} The logarithm of $T_c(L)-T_c$, for
systems of linear size $L=8,16,32,64$, is plotted versus the
logarithm of $L$. The slope, $a$, is $-1/\nu$ and $T_c$ is tuned
to $3.77$ to make the slope of the linear fit to fit the negative
inverse value of $\nu=1.64$.}
\end{figure}

As a consistency check, we use the set of $T_c(L)$'s of the
derivative (as they are better defined than those of the
susceptibility itself), together with our estimation of
$\nu=1.64$, to extract the critical temperature $T_c$, directly
from the finite size scaling behavior,
\begin{equation}
|T_c(L)-T_c| \sim L^{-1/\nu}.
\label{EqTcLFinSizScal}
\end{equation}
In Fig. \ref{FigFitTcL} we use a linear fit for the logarithm of
$T_c(L)-T_c$ plotted versus the logarithm of $L$. We obtain
$T_c=3.77$ and that is by tuning $T_c$ to reach a slope that fits
the negative inverse value of $\nu=1.64$.

\begin{figure}[p]
\includegraphics[width=.88\textwidth]{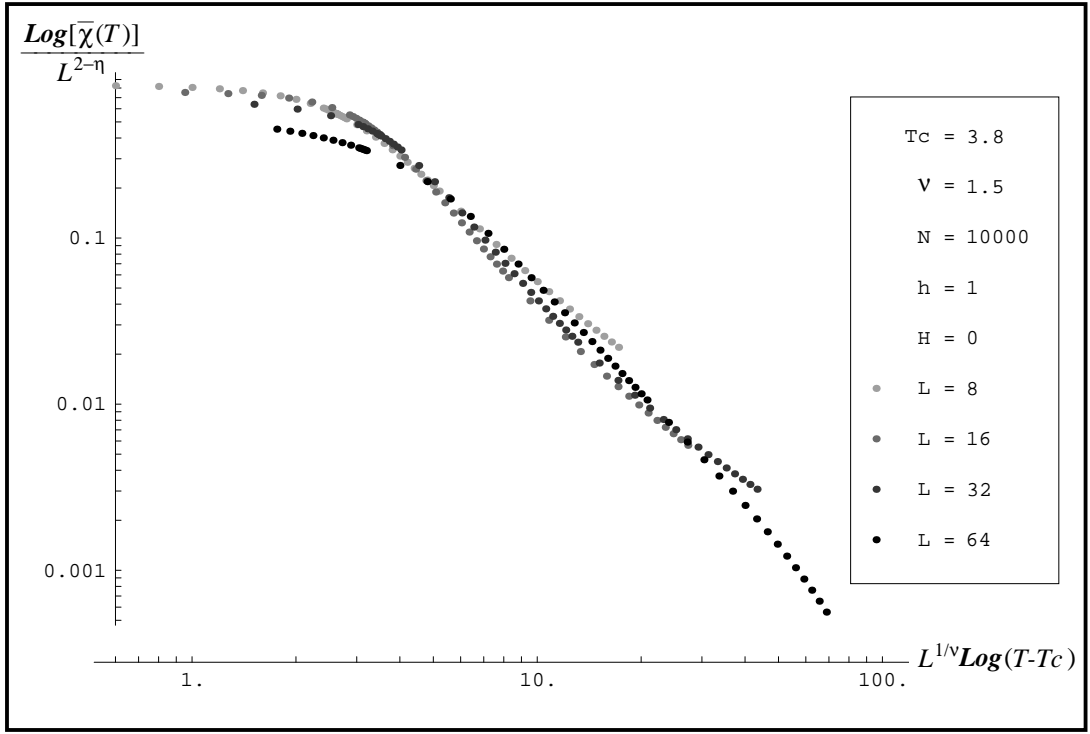}
\caption{\label{FigDataColaps} The logarithm of the
susceptibility, scaled by a factor of $1/L^{2-\eta}$, is plotted
versus the logarithm of $T-T_c$, scaled by a factor of
$L^{-1/\nu}$. The data collapse is shown here for $\eta=0.53$ and
$\nu=1.5$, which, by Eq. (\ref{EqScalRelat}), corresponds to
$\gamma=2.2$. The critical temperature is taken to be $T_c=3.8$.}
\end{figure}

Taking all the results above into account, we arrive at our final
estimation of, $\eta=0.53\pm 0.003$, $\gamma=2.2\pm 0.3$,
$\nu=1.5\pm 0.15$ and $T_c=3.8\pm 0.1$. Note that our result for
$\eta$ satisfies the inequality $2-\eta<d/2$, by Schwartz and
Soffer \cite{ss85} and is in good agreement with the results cited
in Refs. \cite{dsy93,c86,oh86,ry93,r95,fh9698,hy01}. Also, our
result for $\gamma$ is in good agreement with the results of Refs.
\cite{dsy93,gaahs9396,s88,ry93,fhbh97}, while our result for $\nu$
is in good agreement with the results of Refs.
\cite{bcsy85,r95,fhbh97,hy01,mf02}. In Fig. \ref{FigDataColaps} we
use these values to obtain a data collapse for the susceptibility,
scaled by a factor of $1/L^{2-\eta}$, when plotted versus $T-T_c$,
scaled by a factor of $L^{-1/\nu}$. It is presented in a
$log$-$log$ plot. It should be noted, though, that, within the
range of values we have obtained for the parameters above, this
data collapse picture, is almost insensitive, so that it is
impossible to prefer one set of parameters over the other.

\begin{figure}[p]
\includegraphics[width=.88\textwidth]{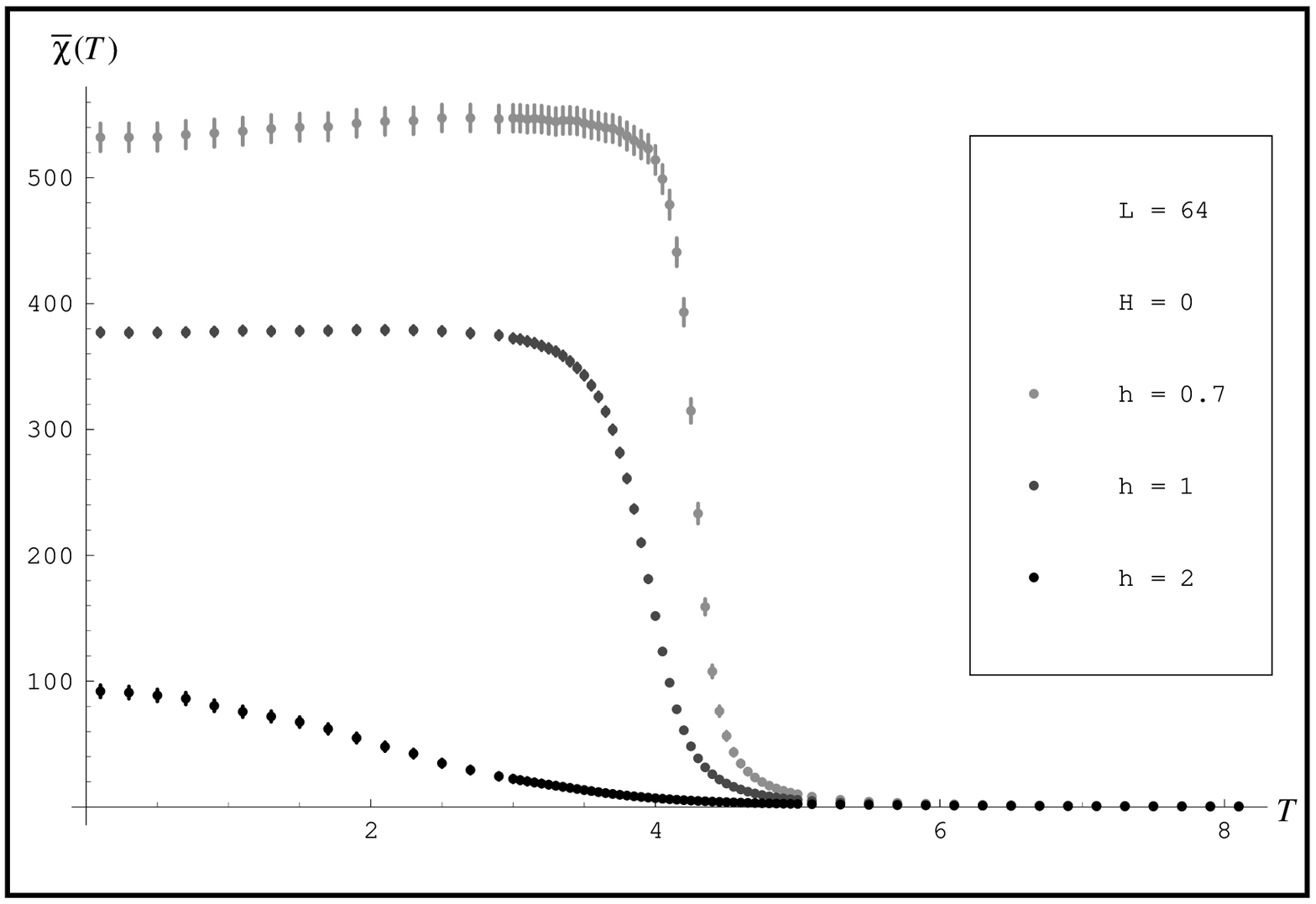}
\caption{\label{FigSusch.7-2} The average susceptibility,
$\overline{\chi}$, is shown as a function of temperature, $T$. The
external field, $H$, is zero, while, as indicated on each figure,
the different levels of grayscaling correspond to different values
of the standard deviation of the random field. The system is of
linear size, $L=64$.}
\end{figure}

We conclude our study of the average susceptibility by presenting
it for different values of the strength of the random field (Fig.
\ref{FigSusch.7-2}).

\begin{figure}[p]
\includegraphics[width=.88\textwidth]{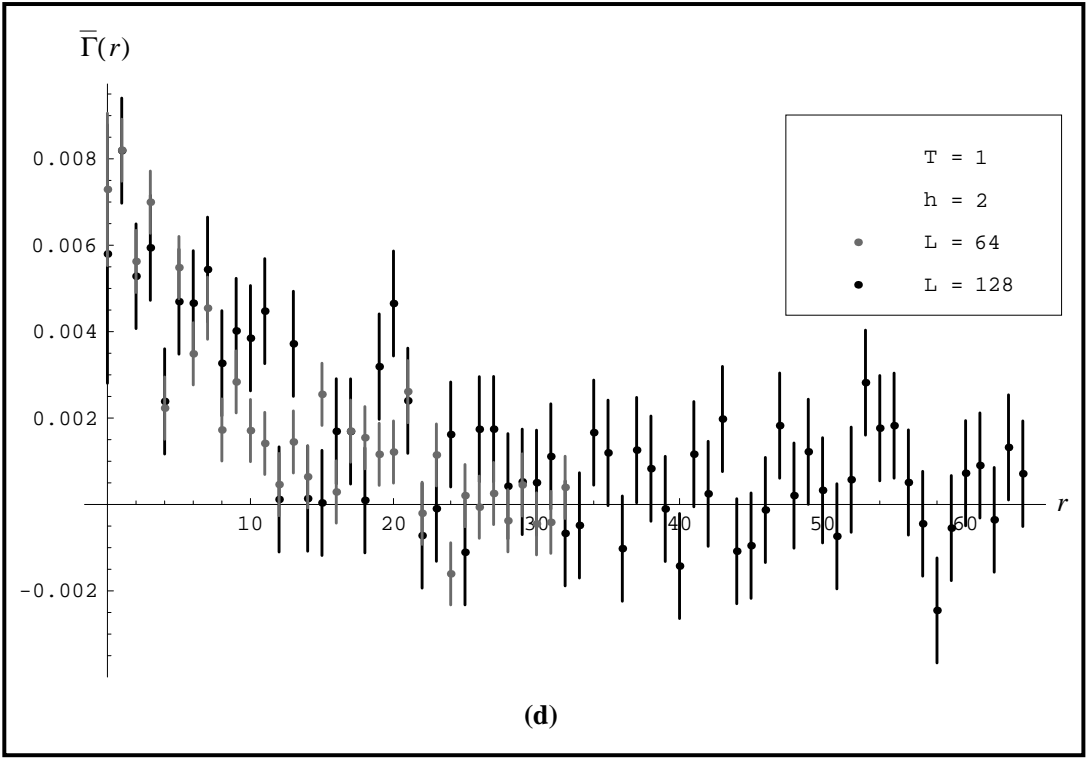}
\caption{\label{FigSScorrL6-7h1-2} The average spin-spin
correlation function, $\overline{\Gamma}$, is shown as a function
of the distance, $r$, taken along the main axes of the lattice and
measured in units of lattice constant. The external field, $H$, is
zero while the standard deviation of the random field is $h=1$ in
(a) and (b) and $h=2$ in (c) and (d). As indicated on each figure,
the two different levels of grayscaling correspond to systems of
different linear size, $L$, indicating the size independence of
$\overline{\Gamma}$ for these temperatures. The points of the
larger system, with $L=128$, appear to be more scattered since it
is averaged only over $3500$ realizations, while the smaller,
$L=64$, system is averaged over $10000$ realizations. The
different figures correspond to different temperatures. Note the
broadening of $\overline{\Gamma}$ as the temperature is reduced
towards entering the ordered phase at about $T=3.8$ for $h=1$ and
at $T\gtrsim 0$ for $h=2$. Also note that for $h=2$, even at a
temperature as low as $T=1$, the correlations are kept relatively
short-ranged indicating the persistence of the disorder phase to
lower temperatures.}
\end{figure}

\begin{figure}[p]
\includegraphics[width=.88\textwidth]{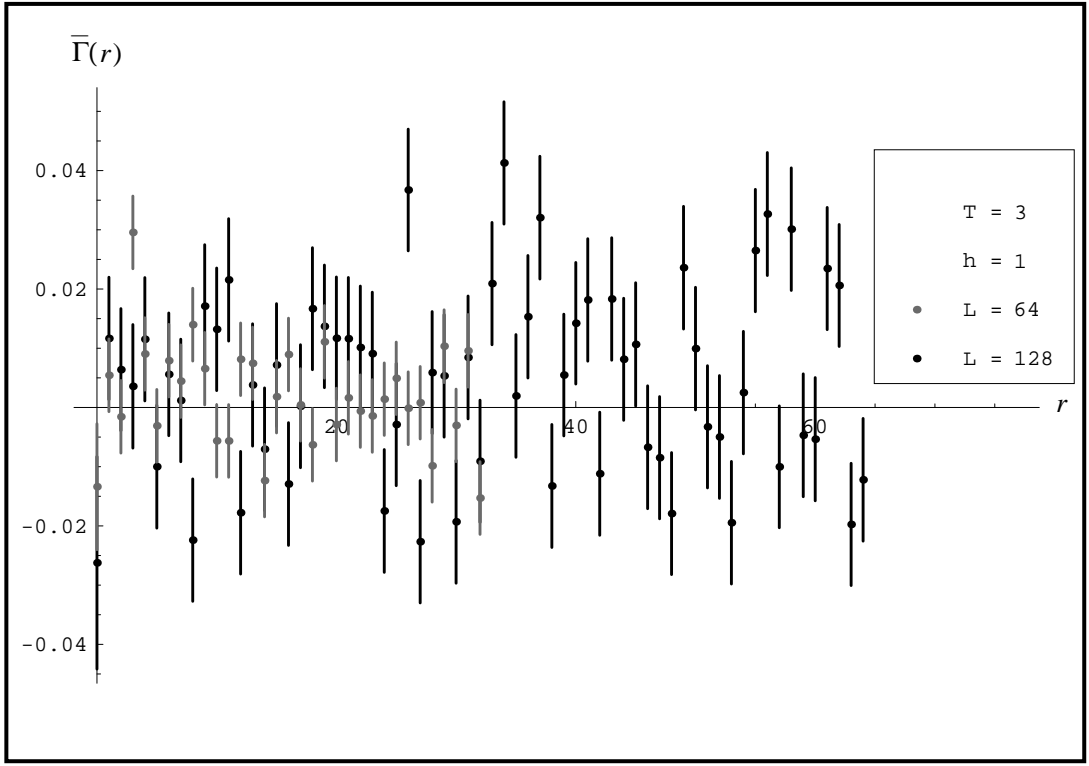}
\caption{\label{FigSScorrL6-7h1T3} The average spin-spin
correlation function, $\overline{\Gamma}$, is again shown as a
function of the distance, $r$, As in Fig. \ref{FigSScorrL6-7h1-2},
only for a point on the phase diagram, $(T,h)=(3.0,1.0)$, located
below the transition. Note the noisiness and lost of trend
compared with points in the phase diagram located above the
transition [Fig. \ref{FigSScorrL6-7h1-2}].}
\end{figure}

\begin{figure}[p]
\includegraphics[width=.88\textwidth]{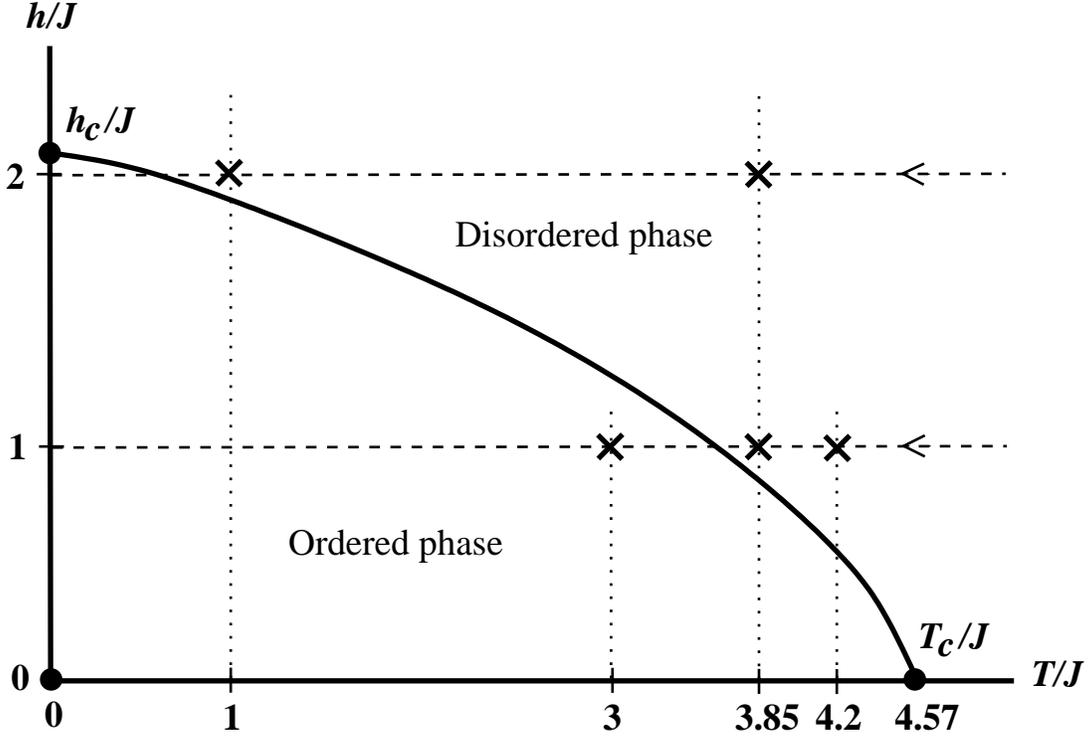}
\caption{\label{FigPhaseDiag} Schematic Phase diagram for the
random field Ising system, are shown for $d=3$ (the thick lines).
The zero temperature fixed point, $(0,h_c/J)$, controls the whole
of the critical line, $h_c(T)$ [or $T_c(h)$], while the zero
fields thermal fixed point, $(T_c/J,0)$, that of the pure Ising
system, is unstable (It is $T_c/J=4.57$ for the Casher-Schwartz
renormalization scheme \cite{cs79}). The lower horizontal dashed
line, represent the lowering of temperature from the high
temperature and disordered phase, at $h=1<h_c$. The critical line
is, thus, crossed and the ordered phase is penetrated. In our
simulations [Fig. \ref{FigSScorrL6-7h1T3}], this is expressed by
the flattening of $\overline{\Gamma}$. The higher horizontal
dashed line, represent the same, only at $h=2\sim h_c$. As
indicated by Fig. \ref{FigSScorrL6-7h1-2}, the level of flattening
of $\overline{\Gamma}$ at a given point $(T/J,h/J)$ on the phase
diagram, depends on the distance of that point from the critical
line. The \textbf{X}'s in the figure represent the points used in
the simulations.}
\end{figure}

We now turn to our evaluation of the average spin-spin correlation
function, $\overline{\Gamma}$, calculated according to Eq.
(\ref{EqTheorem1SSNum}) as discussed in Sec. \ref{SecMethod}. In
Fig. \ref{FigSScorrL6-7h1-2}, we present $\overline{\Gamma(r)}$
for two values of the strength of the random field and for two
temperatures. For $h=1$, we know that $T=4.2$ is above the
transition [Fig. \ref{FigSScorrL6-7h1-2}(a)]. We present our full
results although it is clear that for $r>10$ the values of
$\overline{\Gamma(r)}$ are dominated by noise and therefore
meaningless. For $h=2$ at $T=3.85$ [Fig.
\ref{FigSScorrL6-7h1-2}(c)], we see a similar picture. Again the
function decays very fast and already below $r=10$, its
significance is questionable. It may be expected that increasing
the number of realizations considerably, may improve the
evaluation of the correlation where it is small. As the
temperature is lowered ,for $h=1$, to $T=3.85$ that is at the
transition region [Fig. \ref{FigSScorrL6-7h1-2}(b)], the behavior
becomes very noisy and statistically meaningless, but still a
trend can be discerned. A similar behavior is observed in [Fig.
\ref{FigSScorrL6-7h1-2}(d)], for $h=2$ and $T=1$. We have chosen
to present Figs. \ref{FigSScorrL6-7h1-2}(b) and
\ref{FigSScorrL6-7h1-2}(d) although, as far as
$\overline{\Gamma(r)}$ is concerned, they are not very
informative. The reason for doing so is that it is known
\cite{dsy93,wd95,ah96,wd98,ahw98} that self averaging is destroyed
below the transition and thus we expect the enhanced noisiness of
those figures to indicate approach to the transition. The
existence of a trend suggests, though, that we may be still above
the transition. Fig. \ref{FigSScorrL6-7h1T3} presents
$\overline{\Gamma(r)}$ for $h=1$ and $T=3$ (below the transition).
The noisiness is larger than in Figs. \ref{FigSScorrL6-7h1-2}(b)
and \ref{FigSScorrL6-7h1-2}(d) and no trend as a function of $r$
can be observed. The above observations, in addition to
information about the zero temperature transition ($h_c=1.956$
according to \cite{fh9698}, while it is $h_c=2.28$ and $h_c=2.27$
according to \cite{hy01} and \cite{mf02} respectively) is
consistent with the qualitative phase diagram presented in Fig.
\ref{FigPhaseDiag}. The above may suggest an alternative method of
identifying the critical temperature by the amount of noise in the
data and by loosing the trend as function of $r$. This line of
investigation is postponed, however, to future work.

\begin{figure}[h]
\includegraphics[width=.88\textwidth]{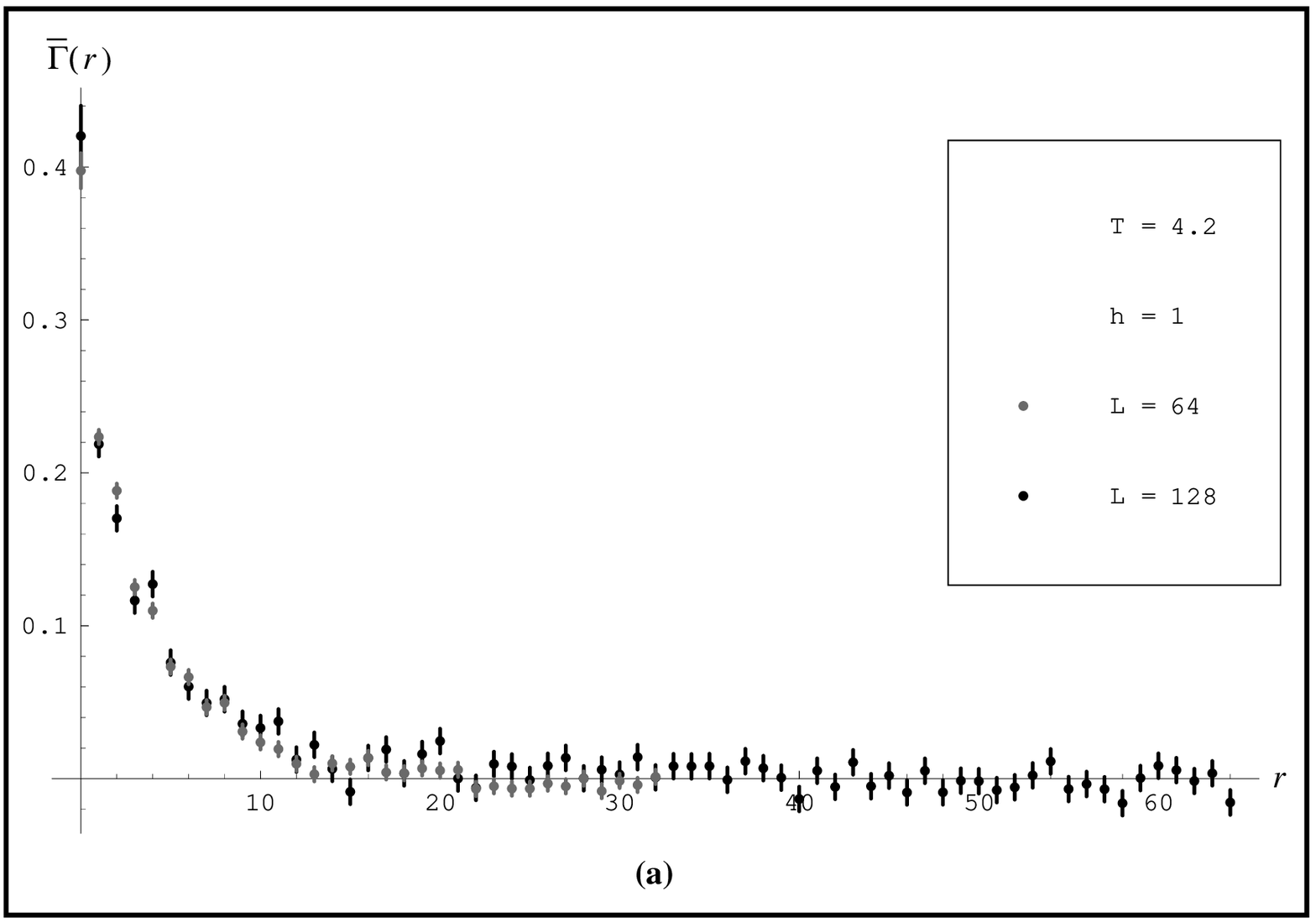}
\includegraphics[width=.88\textwidth]{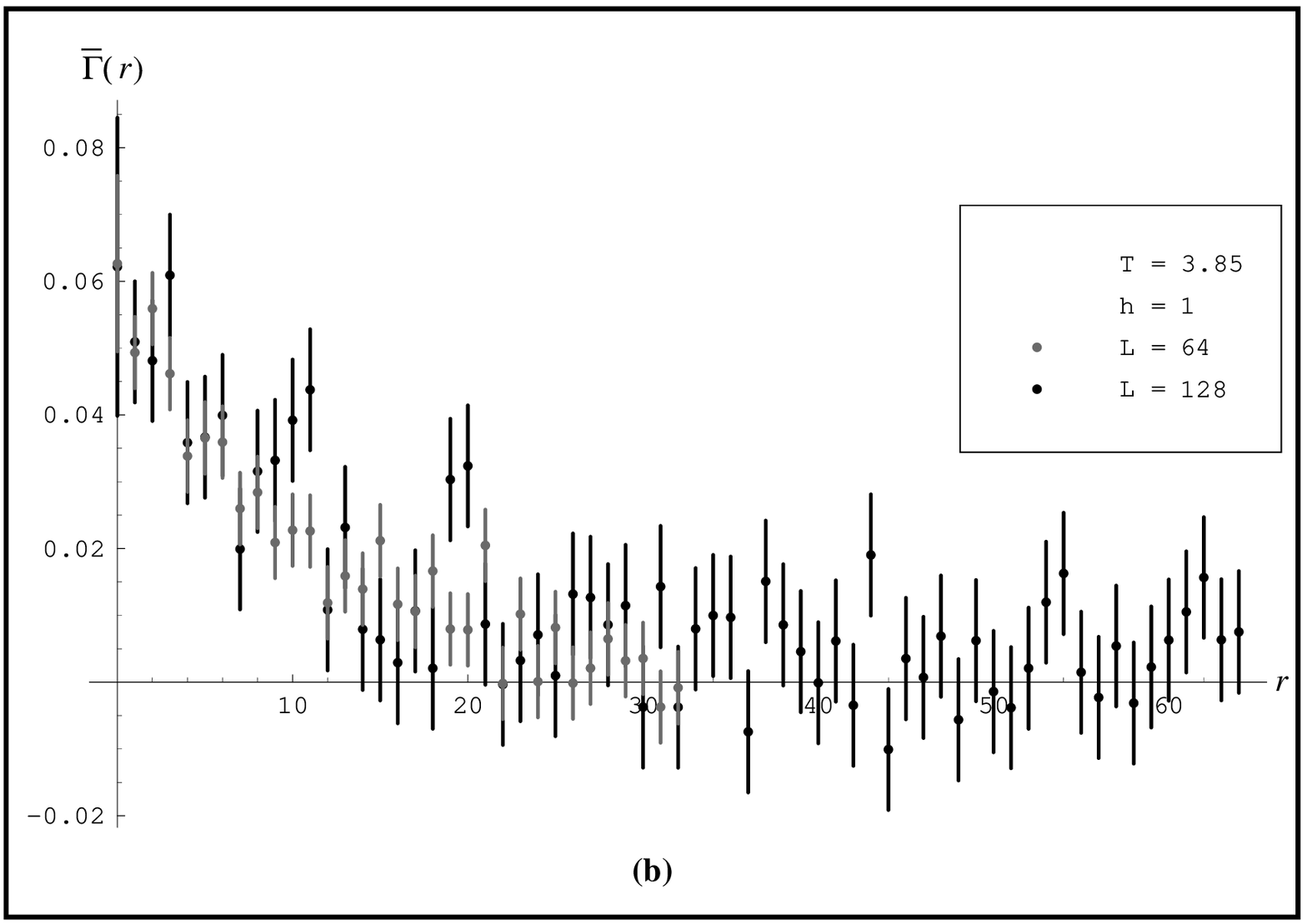}
\end{figure}
\begin{figure}[p]
\includegraphics[width=.88\textwidth]{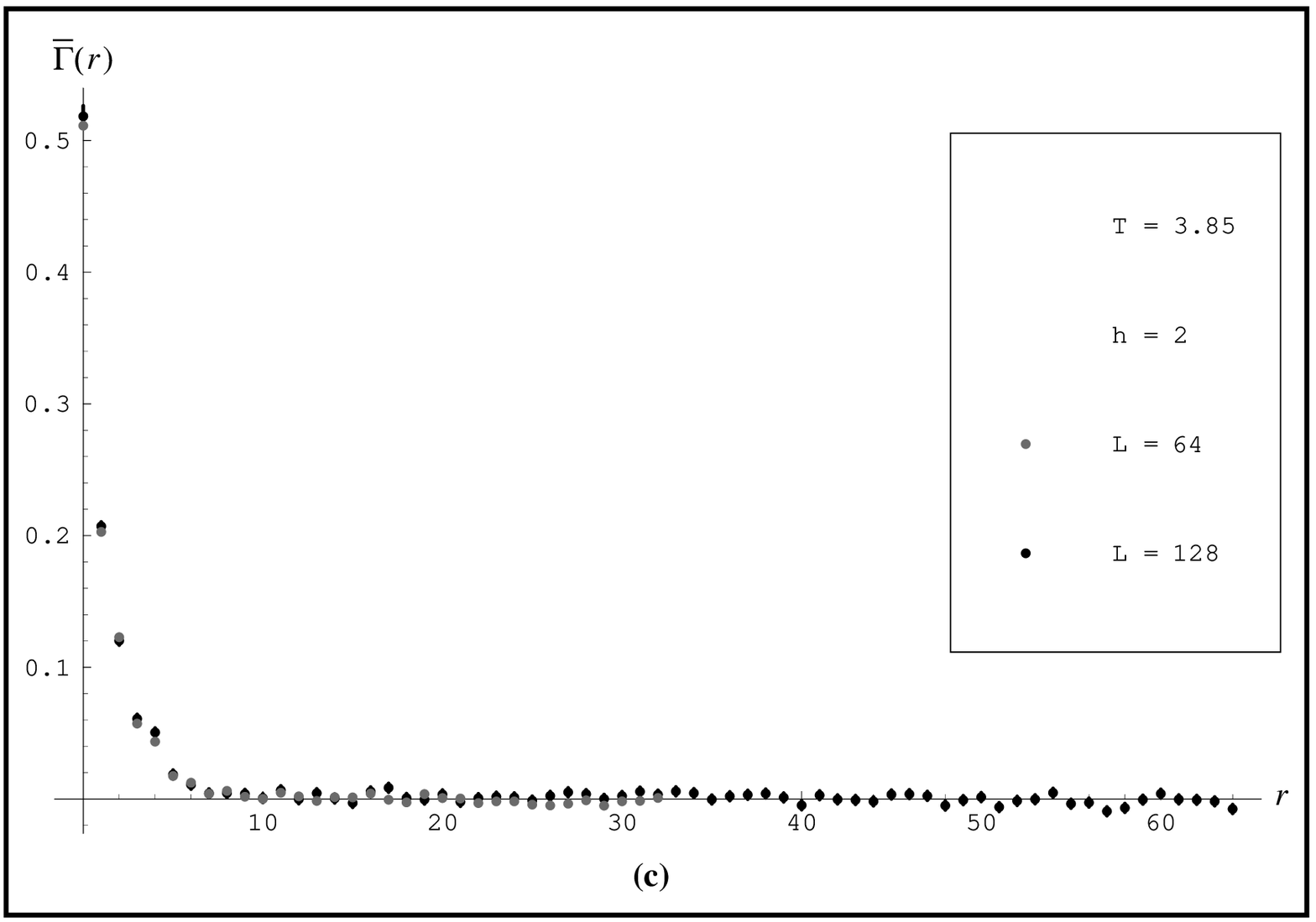}
\end{figure}

\newpage
\section{\label{SecSummary}Summary}
We have presented a method for calculating thermodynamic
quantities, directly from a fully reduced renormalized random
system. Our method works with any renormalization scheme, though
it is essential that the random variables are distributed
according to a Gaussian distribution. It relies on an exact
mathematical transformation, so that the quality of the results
obtained by using it, depends solely on the quality of the
renormalization approximation and the number of realizations
considered. As examples, we have developed explicit expressions
for the average "connected" spin-spin correlations, the average
susceptibility, the average spin-spin correlations
("disconnected"), the total average energy and the average
specific heat. We have demonstrated our method by calculating the
susceptibility and the "connected" correlation function for the 3D
random field Ising system. From the results for the
susceptibility, we have calculated the following critical
exponents: $\eta=0.53\pm 0.003$, $\gamma=2.2\pm 0.3$ and
$\nu=1.5\pm 0.15$ while the critical temperature obtained is
$T_c=3.8\pm 0.1$ for the case where the variance of the field is
$h=1$. As for the average "connected" spin-spin correlation
function, we have presented it as a function of the distance
between spins. Starting at the high temperature and disordered
phase, it shows a sharp decay. Lowering the temperature towards
the critical line it decays over longer and longer distances until
the behavior becomes very noisy and no trend can be detected.



\end{document}